\documentclass[twocolumn]{IEEEtran}
\usepackage{graphicx} 
\usepackage{titling}
\usepackage{setspace}
\usepackage{authblk} 
\usepackage{tcolorbox}
\usepackage{url}
\usepackage{balance}

\title{Towards Human-Bot Collaborative Software Architecting with ChatGPT}
\author[1]{Aakash Ahmad}
\author[2]{Muhammad Waseem}
\author[3]{Peng Liang}
\author[4]{Mahdi Fehmideh}
\author[3]{Mst Shamima Aktar}
\author[2]{Tommi Mikkonen}

\affil[1]{\normalsize School of Computing and Communications, Lancaster University Leipzig, Leipzig, Germany}
\affil[2]{\normalsize Faculty of Information Technology, University of Jyväskylä, Jyväskylä, Finland}
\affil[3]{\normalsize School of Computer Science, Wuhan University, Wuhan, China}
\affil[4]{\normalsize School of Business, University of Southern Queensland, Queensland, Australia}

\affil[ ]{\textsuperscript{}\texttt{a.ahmad13@lancaster.ac.uk}, \texttt{mwaseem@jyu.fi}, \texttt{liangp@whu.edu.cn}, \texttt{mahdi.fahmideh@usq.edu.au}, \texttt{shamima@whu.edu.cn}, \texttt{tommi.j.mikkonen@jyu.fi}}
\date{} 
\begin{document}
\maketitle

\begin{abstract}
Architecting software-intensive systems can be a complex process. It deals with the daunting tasks of unifying stakeholders’ perspectives, designers’ intellect, tool-based automation, pattern-driven reuse, and so on, to sketch a blueprint that guides software implementation and evaluation. Despite its benefits, architecture-centric software engineering (ACSE) inherits a multitude of challenges. ACSE challenges could stem from a lack of standardized processes, socio-technical limitations, and scarcity of human expertise etc. that can impede the development of existing and emergent classes of software (e.g., IoTs, blockchain, quantum systems). Software Development Bots (DevBots) trained on large language models can help synergise architects’ knowledge with artificially intelligent decision support to enable rapid architecting in a human-bot collaborative ACSE. An emerging solution to enable this collaboration is ChatGPT, a disruptive technology not primarily introduced for software engineering, but is capable of articulating and refining architectural artifacts based on natural language processing. We detail a case study that involves collaboration between a novice software architect and ChatGPT for architectural analysis, synthesis, and evaluation of a services-driven software application. Preliminary results indicate that ChatGPT can mimic an architect’s role to support and often lead ACSE, however; it requires human oversight and decision support for collaborative architecting. Future research focuses on harnessing empirical evidence about architects' productivity and exploring socio-technical aspects of architecting with ChatGPT to tackle emerging and futuristic challenges of ACSE.
\end{abstract}


\section{Introduction}
Architecture of software-intensive systems enables architects to specify structural composition, express behavioural constraints, and rationalise design decisions - hiding implementation complexities with architectural components - to sketch a blue-print for software implementation \cite{1_SA200}. Architecture-centric Software Engineering (ACSE) aims to exploit architectural knowledge (e.g., tactics and patterns), architectural languages, tools, and architects’ decisions (human intellect) etc. to create a model that drives the implementation, validation, and maintenance phases of software systems \cite{2_SAProcess2007}. 
In recent years, ACSE has been applied to investigate the role of architecture in engineering complex and emergent classes of software (blockchains, quantum systems etc.) \cite{3_QSA2022} and it has been proven as useful to systematise software development in an industrial context \cite{2_SAProcess2007}. Despite its potential, ACSE entails a multitude of challenges including but not limited to mapping stakeholders’ perspectives to architectural requirements, managing architectural drift, erosion, and technical debts, or lack of automation and architects’ expertise in developing complex and emergent classes of software \cite{1_SA200, 3_QSA2022}. In such context, software engineers may enter a phase referred to as the \textit{`lonesome architect'} who requires non-intrusive support rooted in processes and tools to address the challenges of ACSE by reusing knowledge and exploiting decision support in the process \cite{4_lonesome2011}. 

\textbf{Context and motivation}: The process to architect software applications and services (a.k.a., ‘architecting process’) unifies a number of architecting activities that support an incremental, process-centric, and systematic approach to apply ACSE in software development endeavours \cite{2_SAProcess2007, 3_QSA2022}. Empiricism remains fundamental to deriving and/or utilising architecting processes that can support activities, such as analysis, synthesis, and evaluation etc. of software architetures \cite{4_lonesome2011}. To enrich the architecting process and empower the role of architects, research and development has focused on incorporating patterns and styles (knowledge), recommender systems (intelligence), and distributed architecting (collaboration) in ACSE process. The role of artificial intelligence (AI) in software engineering (SE) is an active area of research that aims to synergise solutions of AI and practices of SE to instill intelligence in the processes and tools for software development \cite{1-xie2018intelligent, 3-barenkamp2020applications}. From an ACSE perspective, research on AI generally aims to develop decision support systems or development bots  that can assist architects with recommendations about design decisions, selection of patterns and styles, or predict points of architectural failure and degradation \cite{5-urli2018design, 4-herold2020towards}. Currently, there is no research that proposes innovative solutions that can enrich the architecting process with AI to enable collaborative architecting. Collaborative architecting can synergise architects' knowledge as human intellect and bot's capability as an intelligent agent who can lead the architecting process based on human conversation and supervision. Such collaboration can allow architects to delegate their architecting tasks to the bot, supervise the bot via dialog in natural language(s) to achieve automation, and relieve architects from undertaking tedious tasks in ACSE.

\textbf{Objective of the study}: Chat Generative Pre-trained Transformer (ChatGPT) has emerged as a disruptive technology, representing an unprecedented example of a bot, that can engage with humans in context-preserved conversations to produce well-articulated responses to complex queries \cite{10-avila2023chatgpt, 9-qadir2022engineering}. ChatGPT is not specifically developed to address software engineering challenges, however; it is well capable of generating versatile textual specifications including architectural requirements, UML scripts, source code libraries, and test cases \cite{8-jalil2023chatgpt, 11-sobania2023analysis}. Recently published research has started to explore the role of ChatGPT in engineering education, software testing, and source code generation \cite{9-qadir2022engineering, 11-sobania2023analysis}. Considering ACSE that can benefit from intelligent and automated architecting, driven by architects’ conversational dialogs and feedback, there is no research to investigate the role that ChatGPT can play as a DevBot in architecting process. To this end, our study focused on a preliminary investigation to understand \textit{if ChatGPT can process an architecture story (scenario(s)) conversed to it by an architect and undertake architecting activities to analyse, synthesise, and evaluate software architecture in a human-bot collaborative architecting}.

\textbf{Contributions}: We followed a process-centric approach \cite{2_SAProcess2007} and adopted scenario-based method \cite{5_SAAM2002} for ChatGPT-enabled architectural analysis, synthesis, and evaluation of a microservices-driven software. Preliminary results demonstrate ChatGPT's capabilities that include but are not limited to processing an architecture story (conversed to it by an architect) for articulating architectural requirements, specifying models, recommending and applying architectural tactics and patterns, and developing scenarios for architecture evaluation. Primary contributions of this study are to: 

\begin{itemize}
    \item Investigate the potential for human-bot collaborative architecting, synergizing ChatGPT’s outputs and architects’ decisions, to automate ACSE with a preliminary case study.
    \item Identify the potential and perils of ChatGPT assisted ACSE to pinpoint issues concerning ethics, governance, and socio-technical constraints of collaborative  architecting.
    \item Establish foundations for empirically-grounded evidence about ChatGPT's capabilities and architects’ productivity in collaborative architecting (ongoing and future work).
\end{itemize}

The results of this study can help academic researchers to formulate new hypotheses about the role of ChatGPT in ACSE and investigate human-bot collaborative architecting of emergent and futuristic software. Practitioners can follow the presented guidelines to experiment with delegating their tedious tasks of ACSE to ChatGPT.

\section{Research Context and Method}
\label{sec:Methodology}
\begin{figure}[ht]
  \begin{center}
  \includegraphics[width=0.55\textwidth]{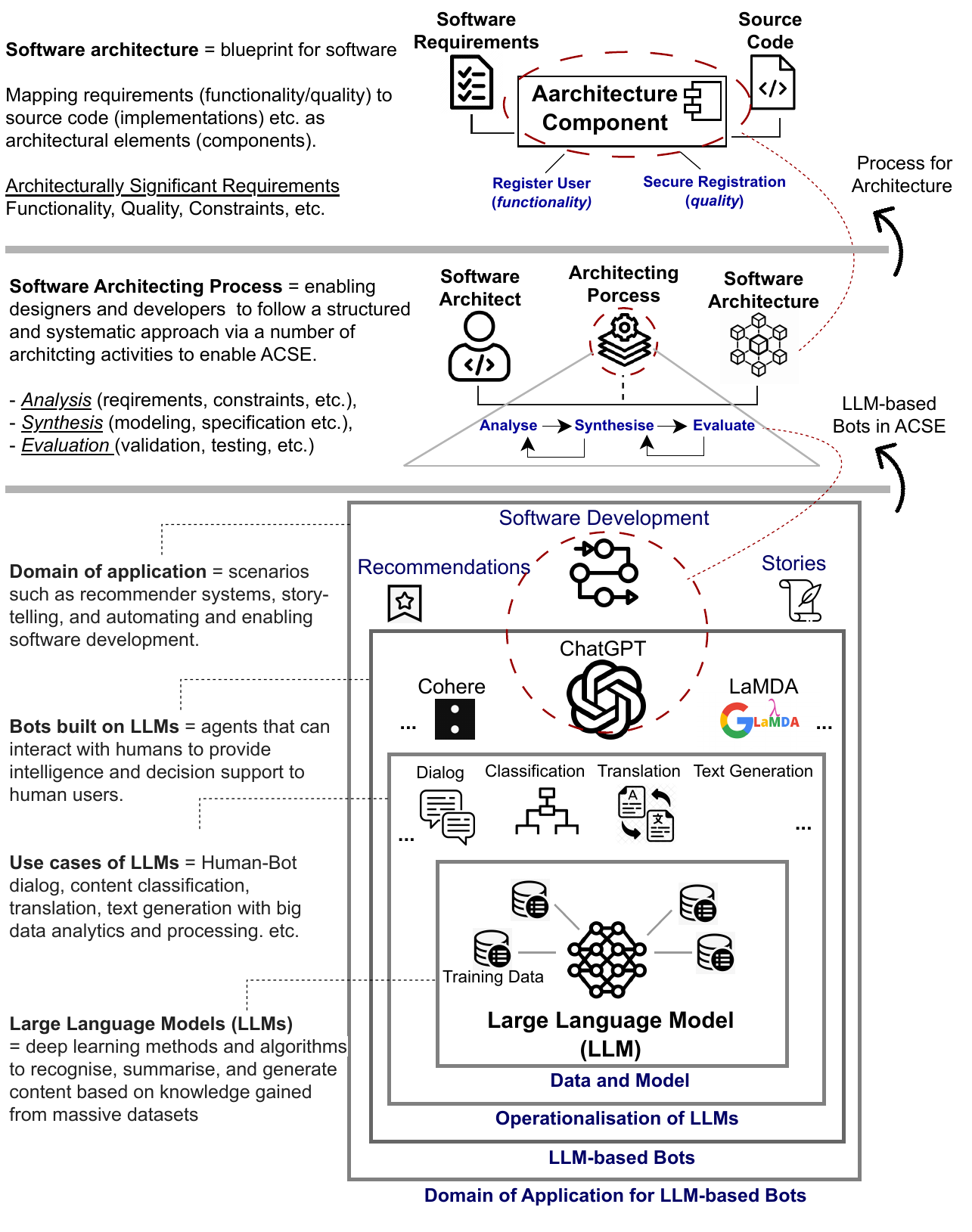}
  \end{center}
  \caption{Context: LLMs, DevBots, Process, and Architecture}
  \label{fig:ResearchContext}
\end{figure}

We next contextualize some core concepts (Section \ref{sec:context}, Figure \ref{fig:ResearchContext}) and discuss the research method (Section \ref{sec:method}, Figure \ref{fig:ResearchMethod}). Terms and concepts introduced here will be used throughout the paper.

\subsection{Human-Bot Collaborative Architecting} \label{sec:context}

\textbf{Software Architecture} as described in the ISO/IEC/IEEE 42010:2011 standard, aims to abstract complexities rooted in source code-based implementations with architectural components and connectors that represent a blueprint of software applications, services, and systems to be developed \cite{1_SA200}. Architecture-centric approaches have proven to be useful in academic solutions as well as in industrial projects by lending architectural knowledge, such as patterns, styles, languages, and frameworks, to design and develop software effectively and efficiently \cite{4_lonesome2011}. To enable software designers and architects with a systematic and incremental design of software architectures, there is a need for \textbf{architecting process} - also referred to as the process for architecting software \cite{2_SAProcess2007, 3_QSA2022}. Architecting process can have a number of fine-grained architecting activities that support a separation of architectural concerns in ACSE. For example, the architecting process reported in \cite{2_SAProcess2007} and illustrated in Figure \ref{fig:ResearchContext} is derived from five industrial projects and incorporates three architecting activities namely \textit{architectural analysis}, \textit{architectural synthesis}, and \textit{architectural evaluation}. For instance, the architectural evaluation activity in the process focuses on scenarios to evaluate the designed architecture \cite{5_SAAM2002}. In the architecting process, an architect can extract and document the requirements that express the required functionality and desired quality of the software, referred to as Architecturally Significant Requirements (ASRs). ASRs need to be mapped to source code implementations via an architectural model that can be visualized or textually specified using architectural languages, such as the Unified Modeling Language (UML) or Architectural Description Languages (ADLs) \cite{6_AL2012}. Architecture models that reflect the ASRs need to be evaluated using an architecture evaluation method, such as Software Architecture Analysis Method (SAAM) or Architecture Tradeoff Analysis Method (ATAM) \cite{5_SAAM2002}.

\textbf{Software Development Bots} (DevBots) represent conversational agents or recommender systems, driven by AI, to assist software engineers by offering certain degree of automation and/or inducing intelligence in software engineering process \cite{5-urli2018design}. From the software architecting perspective, the role of AI in general and DevBots to be specific is limited to bots answering questions or providing recommendations about architectural erosion and maintenance \cite{4-herold2020towards}. There is no research that investigates or any solution that demonstrates an architecting process by incorporating DevBots to enable human-bot collaborative architecting of software systems. Such a collaboration can enrich the architecting process that goes beyond questions \& answers and recommendations, and synergizes architects’ intellect (human rationale) and bot's intelligence (automated architecting process) in ACSE. Collaborative architecting can empower novice designers or architects, who lack experience or professional expertise to specify their requirements in natural language and DevBots can translate them into ASRs, architectural models, and evaluation scenarios. As illustrated in Figure \ref{fig:ResearchContext}, the emergence of ChatGPT as a conversational bot, based on large language models (LLM), can dialog with the architect to lead the creation of architectural artifacts with human supervision. 

\subsection{Research Method}  \label{sec:method}
\begin{figure}[ht]
  \begin{center}
  \includegraphics[width=0.50\textwidth]{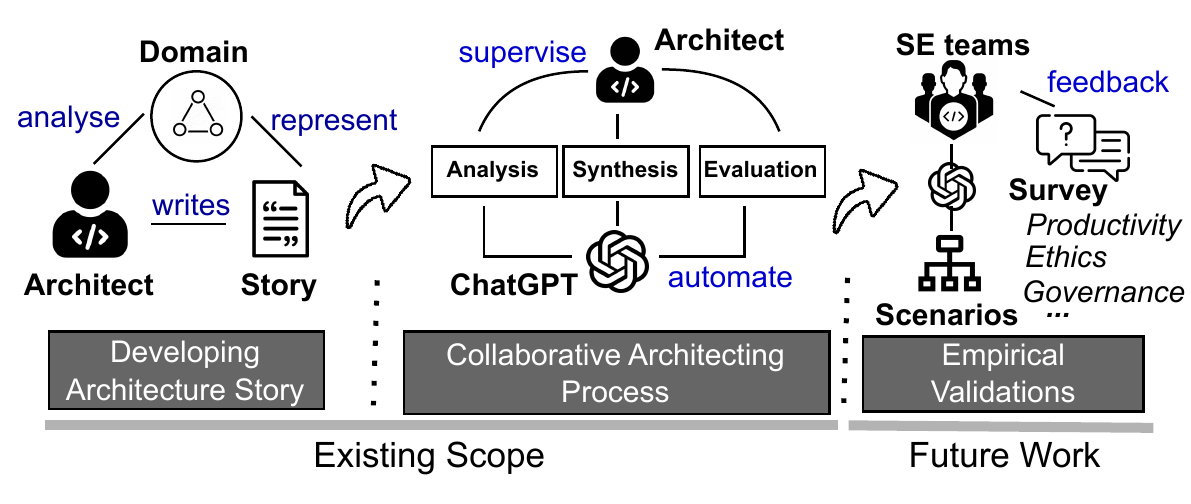}
  \end{center}
  \caption{Overview of the Research Method}
  \label{fig:ResearchMethod}
\end{figure}

We now present the overall methodology for the research, comprising of three main phases, as illustrated in Figure \ref{fig:ResearchMethod}.

\textbf{Phase 1 - Developing the Architecture Story}:
Software architecture story refers to a textual narration of the envisaged solution, i.e., software to be developed by expressing the core functionality, desired quality (i.e., ASRs) and any constraints in a natural language. The story is developed based on analyzing software domain that represents an operational context of the system or collection of scenarios operationalised via a software solution. The architect can analyze the domain and identify scenarios to write an architecture story that acts as a foundation for the architecting process. The architecture story is fed to ChatGPT via a prompt as a pre-process to collaborative architecting. 

\textbf{Phase 2 - Enabling Collaborative Architecting} is 
based on three activities adopted from \cite{2_SAProcess2007}, detailed below.
\begin{itemize}
    \item \textit{Architectural analysis} is driven by architecture story fed to ChatGPT for articulating the ASRs via (i) automatically generated and recommended requirements (by ChatGPT), or (ii) manual specification of the requirements (by the architect), or (iii) a continuous dialog between ChatGPT and the architect to refine (add/remove/update) the requirements. 
    \item \textit{Architectural synthesis} consolidates the ASRs to create an architecture model or representation that can act as a point of reference, visualizing the structural (de-)composition and runtime scenarios for the software. We preferred UML for architectural synthesis due to a number of factors, such as available documentation, ease of use, diversity of diagrams, tool support, and wide-scale adoption as a language to represent software systems \cite{6_AL2012}. During synthesis we also incorporated reuse knowledge and best practices in the form of tactics and patterns to refine the architecture. 
    \item \textit{Architectural evaluation} evaluates the synthesized architecture against ASRs based on scenarios from the architectural story. Architectural evaluation is conducted incrementally for full or partial validation of the architecture or its parts based on use cases or scenarios from ASRs. We used the Software Architecture Analysis Method (SAAM) to supervise ChatGPT for evaluating the architecture \cite{5_SAAM2002}. 
\end{itemize}
\textbf{Phase 3 - Conducting the Empirical Validations} complements the initial two phases with empirical validations of collaborative architecting as an extension of this study, outlining future work. The existing scope aims to explore and present the role of ChatGPT in human-bot collaborative software architecting (in Section \ref{sec:CaseStudy}). Future work on empirically grounded guidelines to understand a multitude of socio-technical issues associated with ChatGPT-driven collaborative architecting is discussed later (in Section \ref{sec:Issues}).

\section{Case Study on Collaborative Architecting} \label{sec:CaseStudy}

\begin{figure}[t]
  \begin{center}
  \includegraphics[width=0.50\textwidth]{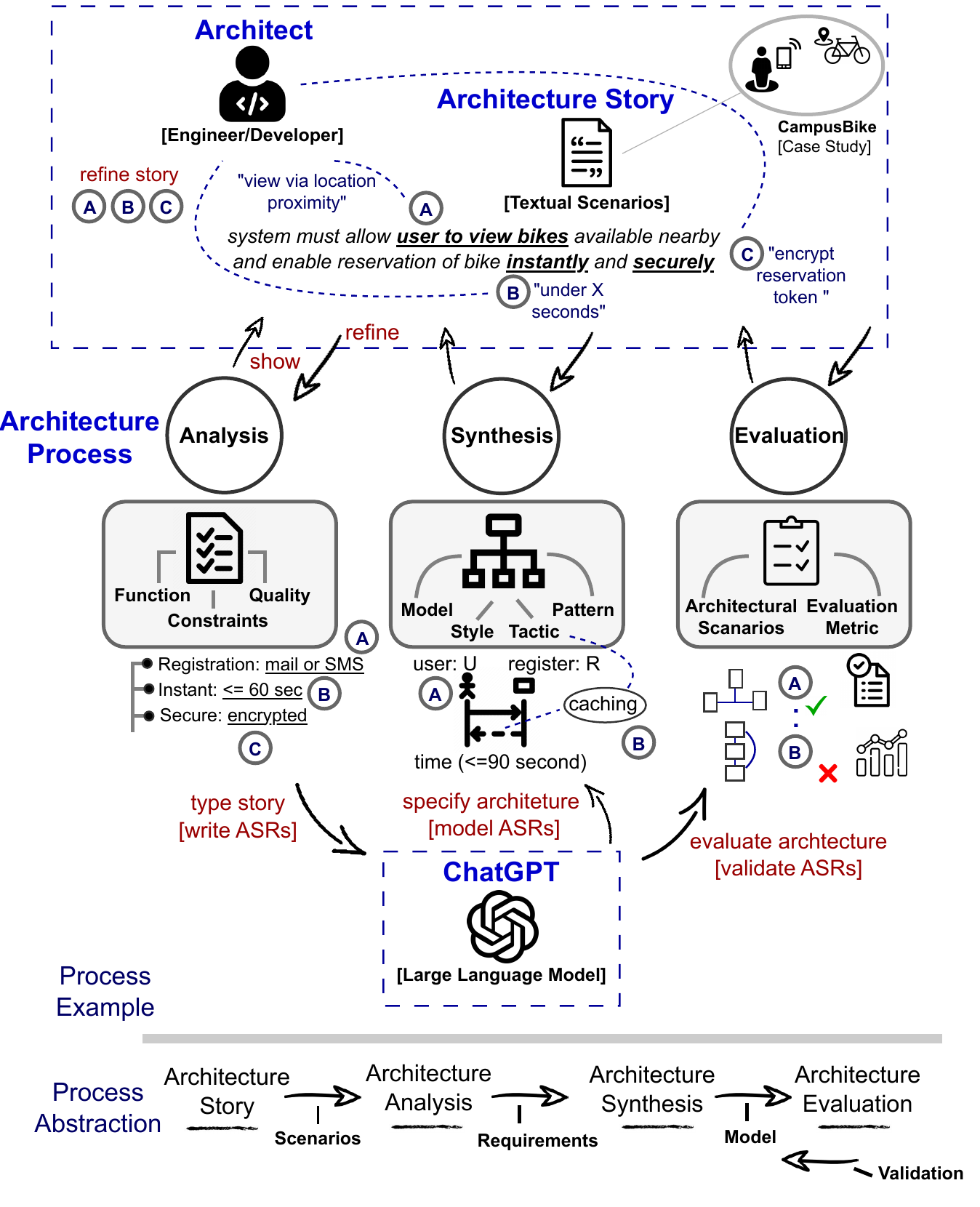}
  \end{center}
  \caption{Overview of the Human-Bot Collaborative Architecting Process}
  \label{fig:SolutionView}
\end{figure}

This section details the process of collaborative architecting demonstrated with a case study for scenario-based exemplification and illustrations (see Figure \ref{fig:SolutionView}). The \textbf{case study} detailed in \cite{Replication} aims to develop a software application named \textsf{CampusBike} that can be used via a browser or as an app, allowing campus visitors to `register', `view available bikes', `reserve a bike', `make payments', and `view usage reports' etc. for eco-friendly mobility in and around the campus. The \textbf{architect} has a working knowledge of software design (UML, patterns etc.) and implementation (programming and scripting languages) and is considered a motivated novice engineer with the responsibility to  design and develop CampusBike software. 

\begin{tcolorbox}
[colback=gray!5!white,colframe=gray!75!black,title=Snippet of Architecture Story]
\scriptsize
`` \ldots~as a step towards maintaining a ‘Green Campus’ - minimising the carbon footprint, congestion, and noise created by vehicles, University’s administration has decided to introduce a bike service where campus visitors can avail of a pay-per-use bicycle facility on an hourly or daily basis for enhanced mobility in and around the campus. Potential bikers can register and view available bikes in their proximity (let us say within 500 meters) and reserve them for a specific time after payment. To facilitate this service, the administration needs a software application called `CampusBike' that is available on Web and mobile devices for potential bikers \ldots''
\vspace{0.2em}

\textbf{Scenario Example}: View available bikes (using location proximity), reserve a bike for a specific time (pay-and-reserve).
\end{tcolorbox}

\subsection{Formulating the Architecture Story}
Architecture story refers to a textual narration of the envisaged solution, i.e., software to be developed by expressing the core functionality and any constraints narrated in a natural language. As per the methodological details in Figure \ref{fig:ResearchMethod}, the story is developed based on analysing software domain that represents an operational context of the system or collection of scenarios operationalized via a software solution. The architect can analyse the domain and identify any scenarios to write an architecture story, fed to ChatGPT, that sets the foundation for architectural analysis activity in the process. Detailed architecture story is available at \cite{Replication}, with its sample snippet and two scenarios highlighted below.

\subsection{Architectural Analysis}
Once the architecture story is fed to ChatGPT, during architectural analysis, the focus is to specify ASRs as required functionality (e.g., view available bikes) and desired quality (e.g., response time < N) along with any constraints (e.g., compliance with relevant data security policies) of CampusBike software. ChatGPT is capable of outlining the ASRs or any necessary constraints if queried by the architect. However, as per the case study, ChatGPT expressed the ASRs and constraints that were refined (add, remove, and modify any requirements) by the architect. For example, the `Reserve Bike' requirement articulated by ChatGPT read as: `\textit{... system must allow user to view bikes available nearby and enable reservation of the bike instantly and securely}’. The architect refined the requirements: 

\begin{tcolorbox}[colback=gray!5, colframe=gray!90, coltext=black]\scriptsize
\textbf{Architect's Refinements} \\Functionality: View Bike - \underline{via location proximity} 
\\Quality: Instantly - \underline{within 90 seconds}, Securely - \underline{encrypt reservation token} \\Constraint: \underline{apply data minimization} on registration data (GDPR constraint)
\end{tcolorbox}

\begin{figure}[t]
  \begin{center}
  \includegraphics[width=0.50\textwidth]{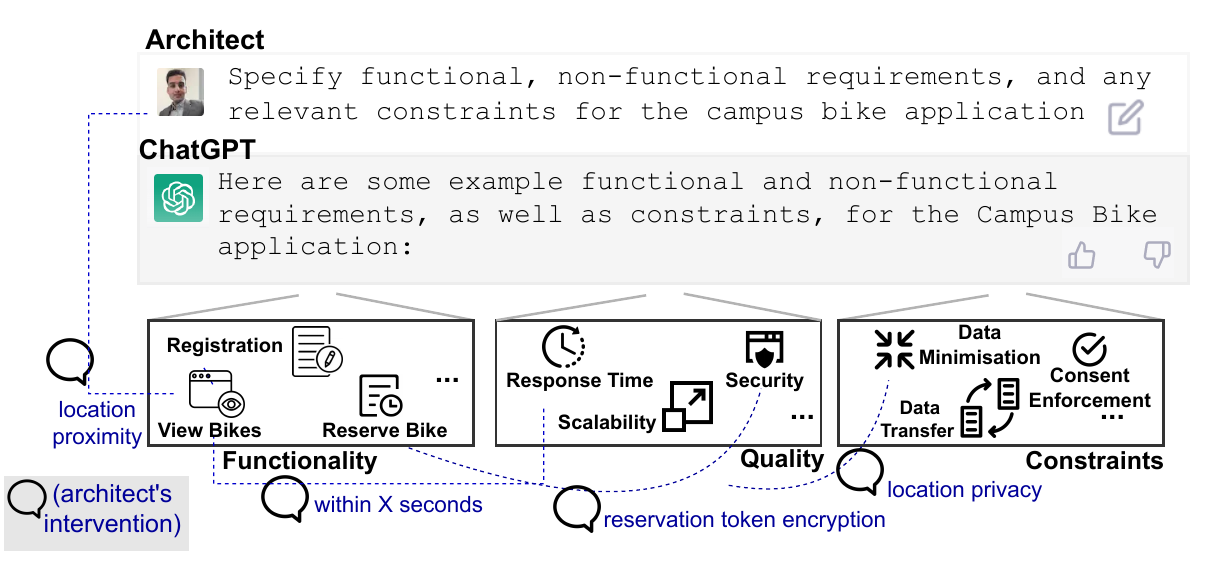}
  \end{center}
  \caption{Formulating and Refining the Requirements}
  \label{fig:analysis}
\end{figure}

After narrating the architecture story, Figure \ref{fig:analysis} shows architects' query and ChatGPT's response (human-bot collaboration) to specify the functionality, quality, and constraints, collectively referred to as the ASRs. ASRs are iteratively refined via a dialog between the two to produce a final list presented here \cite{Replication}.

\subsection{Architectural Synthesis}
The ASRs are synthesized into an architectural model that can be expressed with an architectural (modeling) language, like UML or other architectural languages \cite{6_AL2012}. We used UML class and component diagrams to create the architecture model, specifically; component diagrams to represent the overall architecture, and class diagram for fine-grained representation of the architectural design. During synthesis, we refined the UML class diagram with the application of singleton pattern to  `UserLogin' class to restrict a single login session across the devices. We applied the caching tactic on `ViewBikes' and data minimization constraint on `User Location'.  
\begin{figure}[ht]
  \begin{center}
  \includegraphics[width=0.50\textwidth]{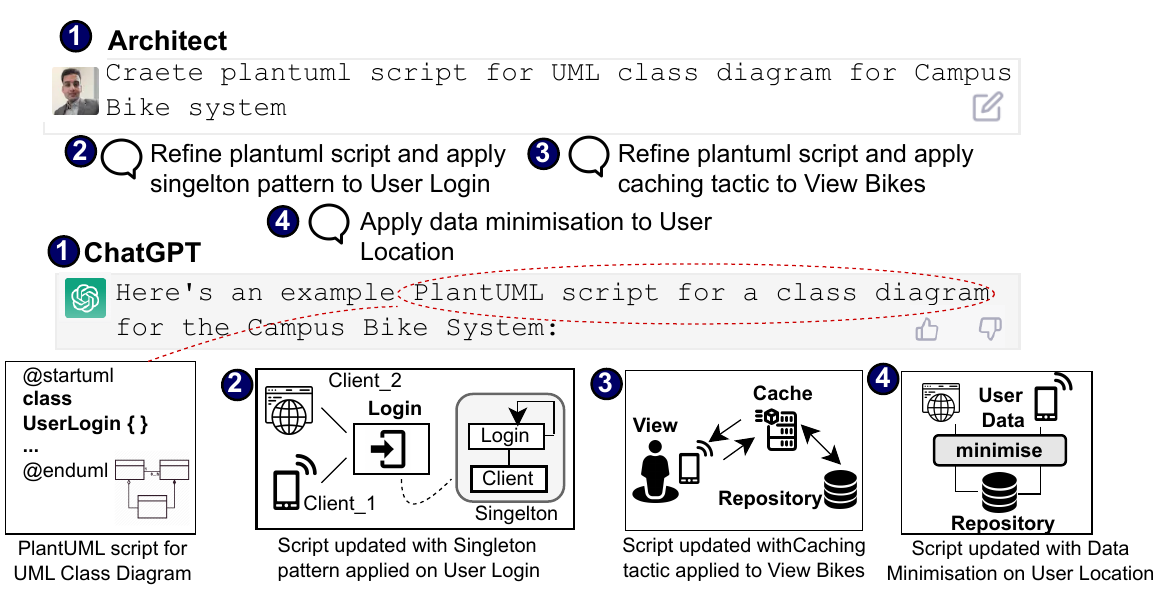}
  \end{center}
  \caption{Modeling and Refining the Architecture Design}
  \label{fig:synthesis}
\end{figure}
 Figure \ref{fig:synthesis} shows the architect's instruction for ChatGPT's to create the script for UML class diagram. Additional dialog between the two enabled application of singleton pattern, caching tactic, and data minimisation constraint on class diagram, presented in \cite{Replication}.
\subsection{Architectural Evaluation}

\begin{figure}[ht]
  \begin{center}
  \includegraphics[width=0.50\textwidth]{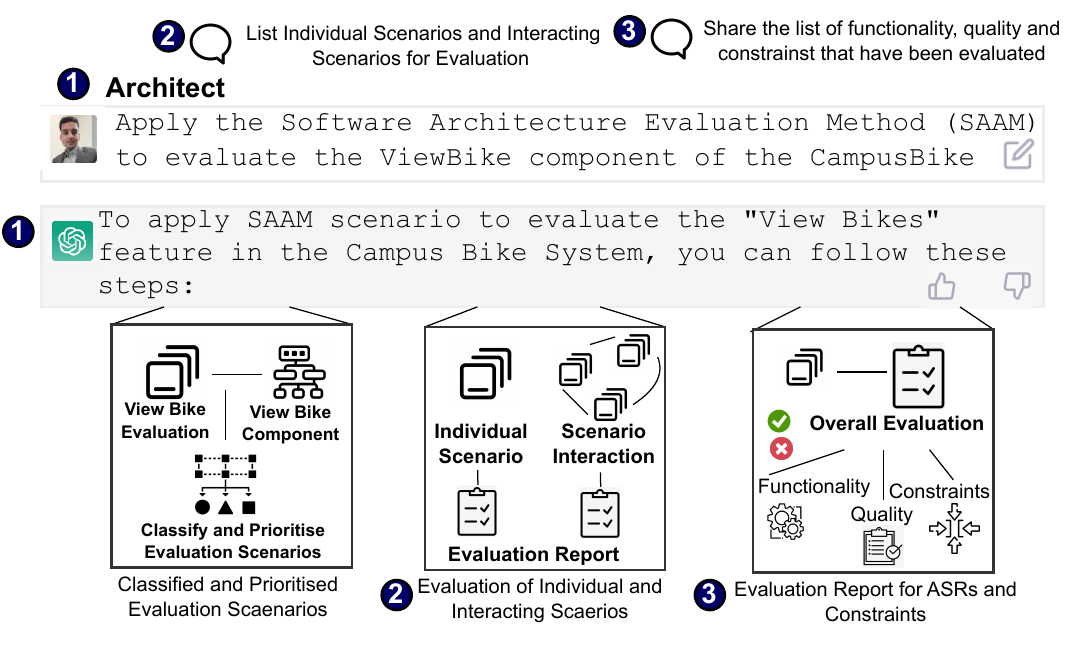}
  \end{center}
  \caption{Evaluating the Architecture}
  \label{fig:evaluation}
\end{figure}

Once synthesized (Figure \ref{fig:analysis}), the architecture needs to be evaluated to assess if it satisfies the ASRs and the constraints (Figure \ref{fig:synthesis}). We have used the SAAM method \cite{5_SAAM2002} to evaluate the synthesized architecture, as illustrated in Figure \ref{fig:evaluation}. For example, the architect specifies the application of SAAM to evaluate the `View Bike' component. ChatGPT presents the scenario for evaluating the `View Bike' component individually and also scenarios where it interacts with other components. Based on the interaction of individual and interacting scenarios, an evaluation report is produced that shows the evaluation of the functionality, quality, and constraints of CampusBike architecture.
\section{Related Work} \label{sec:RelatedWork}
We discuss the most relevant existing research that overviews the application of AI in SE and ACSE (Section \ref{sec:AIinSoftware Engineering}), and the role of ChatGPT in software development (Section \ref{sec:ChatGPTAssisted}). 

\subsection{AI in Software Engineering and Architecting} \label{sec:AIinSoftware Engineering}
The research on synergizing AI and SE can be classified into two distinct dimensions namely AI for SE (artificial intelligence in software engineering) and SE for AI (software engineering for artificial intelligence) \cite{1-xie2018intelligent} \cite{3-barenkamp2020applications}. Considering the AI for SE perspective, Xie \cite{1-xie2018intelligent} argued that SE research needs to go beyond traditional efforts of applying AI for tool-based automation and pattern selection with an exploration of methods that instil intelligence in software engineering processes and solutions. Specifically, SE solutions need to maintain the so-called `intelligence equilibrium' – i.e., unifying and balancing machine intelligence and human intellect – in processes, patterns, and tools etc. for emergent classes of software, such as blockchain and quantum applications \cite{2-woods2016software}. Barenkamp \textit{et al}. \cite{3-barenkamp2020applications} combined the findings of a systematic review and interviews with software developers to investigate the role of AI techniques in SE processes. The results of their study pinpoint three areas where SE needs intelligence to tackle (i) automation of tedious and complex SE activities such as code debugging, (ii) big data analytics to discover patterns, and (iii) evaluation of data in neural and software-defined networks. Considering the context of AI in software architecting, Herold \textit{et al}. \cite{4-herold2020towards} investigated existing research and proposed a conceptual framework for the application of machine learning to mitigate architecture degradation. 

\subsection{ChatGPT Assisted Software Engineering} \label{sec:ChatGPTAssisted}
From the SE perspective, ChatGPT is viewed as an unprecedented example of a chatbot that can produce well-articulated responses to complex queries. However, it remains an unexplored territory in terms of its potential and perils in the context of software development processes \cite{6-borji2023categorical, 7-doglio_2022}. Most recently, a number of proposals and experimental findings indicate that the research on ChatGPT focuses on supporting engineering education \cite{8-jalil2023chatgpt, 9-qadir2022engineering}, software programming \cite{7-doglio_2022, 10-avila2023chatgpt}, and testing \cite{11-sobania2023analysis}. Avila-Chauvet \textit{et al}. \cite{10-avila2023chatgpt} detailed how conversational dialogs of a programmer with ChatGPT enable a human-bot assisted development of an online behavioral task using HTML, CSS, and JavaScript source code. They highlighted that although ChatGPT requires human oversight and intervention, it can write well-scripted programming solutions and reduces the time and effort of a developer during programming. A similar narrative in a blogpost \cite{7-doglio_2022} advocated for an incremental process (human dialog with ChatGPT) to enable genetic programming - JavaScript code to solve the traveling salesman problem. In addition to developing the source code, a couple of studies have focused on testing and debugging with ChatGPT \cite{8-jalil2023chatgpt, 11-sobania2023analysis}. Sobania \textit{et al}. \cite{11-sobania2023analysis} evaluated the performance of ChatGPT in automated bug fixing. In contrast to the status-quo on automated techniques for bug fixing \cite{5-urli2018design}, ChatGPT offers a dialogue with a software tester for an incremental identification and fixing of bugs.

\textbf{Conclusive summary}: Based on a review of the existing literature, there do not exist any research or development that explores the role of ChatGPT (LLM-driven AI) that can engage software engineers in conversational dialogs to lead and support ACSE. This study complements the most recent research efforts on software test automation and bug fixing with ChatGPT \cite{11-sobania2023analysis} but focuses on architecture-centric development for software systems. In the broader context of AI for SE \cite{1-xie2018intelligent}, this study argues for human-bot collaborative architecting that can enrich ACSE process with the architects' knowledge and supervision synergized with bot's capabilities to architect software-intensive systems and services.  

\section{Discussion and Validity Threats} \label{sec:Issues}
We discuss the socio-technical aspects of collaborative architecting (Section \ref{sociotech}) and highlight potential threats to validity (Section \ref{threats}).

\subsection{Socio-Technical Issues of ChatGPT in ACSE} \label{sociotech}
In addition to highlighting ChatGPT’s potential, we also highlight some perils as shortcomings of collaborative architecting process that need to be discussed in the context of socio-technical aspects. By socio-technical aspects, we refer to a unified perspective on issues such as \textit{what can be `social' concerns} and \textit{what are the `technical' limitations} of collaborative architecting. Dedicated research is required to systematically investigate such issues, however, we only pinpoint several prominent ones, as below. 

\textbf{Response Variation}: In the context of human-bot conversational dialogs, ChatGPT may produce varied responses for exact same queries. For example, we observed that a query such as `\textit{... what architectural style can be best suited to CampusBike system}' may yield varied responses, such as microservices, layered, client-server etc. architecture can be best suited for the system. This and alike variation in recommendations or scripted artifacts (UML script, ASR specification etc.) can impact the consistency of architecting process and ultimately varied analysis, synthesis, and evaluation of the architecture. One of the solutions to minimize response variations is an iterative dialog with ChatGPT to refine its output and architects’ oversight to ensure that the architectural artifacts being produced are consistent and coherent.
    
\textbf{Ethics and Intellectual Property}: Textual specifications, architecture specific scripts, and source codes etc. articulated by ChatGPT could give rise to ethical issues or in some cases copyright or intellectual property infringements. For example, ChatGPT generated script for a component (\textsc{getLocation}) that senses user location in CampusBike system may lead to leakage of user location privacy and non-compliant software with regulatory guidelines (GDPR, CCPA etc.) that must be dealt with vigilance. In such cases, the role of architect is critical to ensure the generated architecture does not violate ethics or intellectual property rights (if any). 
    
\textbf{Biased Outputs}: The biases in outputs of such conversational bots can be attributed to a number of possible aspects including but not limited to input, training data, and/or algorithmic bias. From an architectural perspective, recommendation bias about specific architectural modeling notation, tactic, pattern, or style etc. may be based on its widespread adoption or bias in training data rather than optimal use in a specific context. Moreover, architectural recommendations (specific style), design decisions (pattern selection), or validation scenarios (evaluation method) may suffer such bias to produce sub-optimal artifacts in ACSE. 

\subsection{Threats to the Validity} \label{threats}
 Validity threats represent limitations, constraints, or potential flaws in the study that can affect the generalization, replicability, and validity of results. Future work can focus on minimizing these threats to ensure methodological rigor and generalization of results. 

\textbf{Internal validity} examines the extent to which any systematic error (bias) is present in the design, conduct, and analysis etc. of the study. To design and conduct this study, and considering the internal validity, we followed a systematic approach and utilized a well-known architecting process \cite{2_SAProcess2007} and architecture evaluation method \cite{5_SAAM2002}. The case study based approach combined with incremental architecting (Figure \ref{fig:SolutionView}) helped us to analyze and refine the study, however, more work is required to understand if the study can be validated with a different architecting process or by adopting other evaluation methods. 

\textbf{External validity} examines whether the findings of a study can be generalized to other contexts. We only experimented with a single case study of moderate complexity that can compromise study’s generalization. Specifically, scenarios with the increased complexity of architecting process (cross-organisational development), class of software to be developed (mission-critical software), and human expertise (novice/experienced engineers) can affect the external validity of this research. Future work is planned, highlighted in the conclusions section, to validate the process of collaborative architecting by engaging architecting teams and analyzing their feedback to understand the extent to which the external validity can be minimized.

\textbf{Conclusion validity} determines the degree to which the conclusions reached by the study are credible or believable. In order to minimize this threat, we followed a three-step process (Figure \ref{fig:ResearchMethod}) to support a fine-grained process to architect the software and validate the results (future work). Moreover, a case study based approach was adopted to ensure scenario-based demonstration of the study results. However, some conclusions (e.g., architect’s productivity, ChatGPT’s efficacy) can only be validated with more experimentation involving multiple case studies, diverse teams, and real context scenarios of collaborative architecting.  

\section{Conclusions and Future Research}
\label{sec:conclusions}
ChatGPT has emerged as a disruptive technology, an unprecedented conversational bot, that mimics human conversations and generates well-articulated textual artifacts (recommendation, scripts, source codes etc.) - often referred to as a `solution that seeks a problem'. Among a plethora of its use cases that range from content creation to digital assistance and acting as a virtual teacher etc., ChatGPT’s role as a DevBot and its capability to architect software-intensive systems remain unexplored. This research investigates the potential and perils of ChatGPT to assist and empower the role of an architect who leads the process of architecting, and collaborate with a human to enable ACSE. The research advocates that in the context of AI for SE, traditional efforts of applying AI for tool-based automation should focus on a broader perspective, i.e., enriching existing processes by instilling intelligence in them via efforts like human-bot collaborative architecting. The case study reflects a practical case of \textit{how a software can be architected with ChatGPT?} and \textit{what factors need to be considered in collaborative architecting?} Variance in responses and artifacts, types of ethical implications, level of human decision support/supervision, along with legal and socio-technical issues must be considered while integrating ChatGPT in SE or ACSE processes. The research needs empirical validations, grounded in evidence and experimentation, to objectively assess factors like enhancing engineers’ productivity, SE process optimization, and assisting novice developers and designers to engineer complex and emergent classes of software effectively with ChatGPT. 

\textbf{Needs for future research}: We plan to extend this study as a stream of research that explores human feedback and validation (i.e., architects’ perspective) and integrating ChatGPT in a process to develop software services for quantum computing systems. More specifically, quantum computing and quantum software engineering has emerged as a quantum computing genre of SE that faces a lack of human expertise to synergize the skills of engineering software and knowledge of quantum physics. We are currently working in engaging a number of software development teams with diverse demography attributes (e.g., geo-distribution, type of expertise, level of experience, class of software system) in controlled experiments to architect software systems using ChatGPT and document architects’ responses. Specifically, with a case study that involves ChatGPT assisted architecting shall allow us to capture feedback of architects via interviews or documents to empirically investigate aspects like usefulness, rigor, acceptance, impact on human productivity, and potential perils of ChatGPT in ACSE.

\bibliographystyle{IEEEtran}
\bibliography{References}

\begin{thebibliography}{10}
\providecommand{\url}[1]{#1}
\csname url@samestyle\endcsname
\providecommand{\newblock}{\relax}
\providecommand{\bibinfo}[2]{#2}
\providecommand{\BIBentrySTDinterwordspacing}{\spaceskip=0pt\relax}
\providecommand{\BIBentryALTinterwordstretchfactor}{4}
\providecommand{\BIBentryALTinterwordspacing}{\spaceskip=\fontdimen2\font plus
\BIBentryALTinterwordstretchfactor\fontdimen3\font minus
  \fontdimen4\font\relax}
\providecommand{\BIBforeignlanguage}[2]{{%
\expandafter\ifx\csname l@#1\endcsname\relax
\typeout{** WARNING: IEEEtran.bst: No hyphenation pattern has been}%
\typeout{** loaded for the language `#1'. Using the pattern for}%
\typeout{** the default language instead.}%
\else
\language=\csname l@#1\endcsname
\fi
#2}}
\providecommand{\BIBdecl}{\relax}
\BIBdecl

\bibitem{1_SA200}
P.~Kruchten, H.~Obbink, and J.~Stafford, ``The past, present, and future for
  software architecture,'' \emph{IEEE Software}, vol.~23, no.~2, pp. 22--30,
  2006.

\bibitem{2_SAProcess2007}
C.~Hofmeister, P.~Kruchten, R.~L. Nord, H.~Obbink, A.~Ran, and P.~America, ``A
  general model of software architecture design derived from five industrial
  approaches,'' \emph{Journal of Systems and Software}, vol.~80, no.~1, pp.
  106--126, 2007.

\bibitem{3_QSA2022}
A.~Ahmad, A.~A. Khan, M.~Waseem, M.~Fahmideh, and T.~Mikkonen, ``Towards
  process centered architecting for quantum software systems,'' in
  \emph{Proceedings of the 1st IEEE International Conference on Quantum
  Software (QSW)}.\hskip 1em plus 0.5em minus 0.4em\relax IEEE, 2022, pp.
  26--31.

\bibitem{4_lonesome2011}
J.~F. Hoorn, R.~Farenhorst, P.~Lago, and H.~Van~Vliet, ``The lonesome
  architect,'' \emph{Journal of Systems and Software}, vol.~84, no.~9, pp.
  1424--1435, 2011.

\bibitem{1-xie2018intelligent}
T.~Xie, ``Intelligent software engineering: Synergy between ai and software
  engineering,'' in \emph{Proceedings of the 11th Innovations in Software
  Engineering Conference (ISEC)}.\hskip 1em plus 0.5em minus 0.4em\relax ACM,
  2018, pp. 1--1.

\bibitem{3-barenkamp2020applications}
M.~Barenkamp, J.~Rebstadt, and O.~Thomas, ``Applications of ai in classical
  software engineering,'' \emph{AI Perspectives}, vol.~2, no.~1, p.~1, 2020.

\bibitem{5-urli2018design}
S.~Urli, Z.~Yu, L.~Seinturier, and M.~Monperrus, ``How to design a program
  repair bot? insights from the repairnator project,'' in \emph{Proceedings of
  the 40th International Conference on Software Engineering: Software
  Engineering in Practice (ICSE-SEIP)}.\hskip 1em plus 0.5em minus 0.4em\relax
  ACM, 2018, pp. 95--104.

\bibitem{4-herold2020towards}
S.~Herold, C.~Knieke, M.~Schindler, and A.~Rausch, ``Towards improving software
  architecture degradation mitigation by machine learning,'' in
  \emph{Proceedings of the 12th International Conference on Adaptive and
  Self-Adaptive Systems and Applications (ADAPTIVE)}.\hskip 1em plus 0.5em
  minus 0.4em\relax IARIA, 2020, pp. 36--39.

\bibitem{10-avila2023chatgpt}
L.~Avila-Chauvet, D.~Mej{\'\i}a, and C.~O. Acosta~Quiroz, ``Chatgpt as a
  support tool for online behavioral task programming,'' \emph{SSRN preprint
  SSRN:4329020}, 2023.

\bibitem{9-qadir2022engineering}
J.~Qadir, ``Engineering education in the era of chatgpt: Promise and pitfalls
  of generative ai for education,'' \emph{TechRxiv preprint techrxiv.21789434},
  2022.

\bibitem{8-jalil2023chatgpt}
S.~Jalil, S.~Rafi, T.~D. LaToza, K.~Moran, and W.~Lam, ``Chatgpt and software
  testing education: Promises \& perils,'' \emph{arXiv preprint
  arXiv:2302.03287}, 2023.

\bibitem{11-sobania2023analysis}
D.~Sobania, M.~Briesch, C.~Hanna, and J.~Petke, ``An analysis of the automatic
  bug fixing performance of chatgpt,'' \emph{arXiv preprint arXiv:2301.08653},
  2023.

\bibitem{5_SAAM2002}
L.~Dobrica and E.~Niemela, ``A survey on software architecture analysis
  methods,'' \emph{IEEE Transactions on Software Engineering}, vol.~28, no.~7,
  pp. 638--653, 2002.

\bibitem{6_AL2012}
I.~Malavolta, P.~Lago, H.~Muccini, P.~Pelliccione, and A.~Tang, ``What industry
  needs from architectural languages: A survey,'' \emph{IEEE Transactions on
  Software Engineering}, vol.~39, no.~6, pp. 869--891, 2012.

\bibitem{Replication}
A.~Ahmad, M.~Waseem, P.~Liang, M.~Fehmideh, M.~S. Aktar, and T.~Mikkonen,
  ``Replication package for the paper: Towards human-bot collaborative software
  architecting with chatgpt.''\hskip 1em plus 0.5em minus 0.4em\relax
  \url{https://github.com/shamimaaktar1/ChatGPT4SA}, 2023.

\bibitem{2-woods2016software}
E.~Woods, ``Software architecture in a changing world,'' \emph{IEEE Software},
  vol.~33, no.~6, pp. 94--97, 2016.

\bibitem{6-borji2023categorical}
A.~Borji, ``A categorical archive of chatgpt failures,'' \emph{arXiv preprint
  arXiv:2302.03494}, 2023.

\bibitem{7-doglio_2022}
\BIBentryALTinterwordspacing
F.~Doglio, ``The rise of chatgpt and the fall of the software developer - is
  this the beginning of the end?'' Dec 2022. [Online]. Available:
  \url{https://tinyurl.com/3mxrfmjh}
\BIBentrySTDinterwordspacing

\end{thebibliography}

\balance
\end{document}